\documentclass[conference]{IEEEtran}
\IEEEoverridecommandlockouts
\usepackage{cite}
\usepackage{amsmath,amssymb,amsfonts}
\usepackage{algorithmic}
\usepackage{graphicx}
\usepackage{textcomp}
\usepackage[table]{xcolor}
\def\BibTeX{{\rm B\kern-.05em{\sc i\kern-.025em b}\kern-.08em
    T\kern-.1667em\lower.7ex\hbox{E}\kern-.125emX}}

\usepackage{multirow}
\usepackage{booktabs}
\usepackage{hyperref}
\usepackage{cleveref}

\crefname{figure}{Fig.}{Figs.}
\crefname{table}{Table}{Tables.}
    
\begin{document}

\title{Achieving Carbon Neutrality for I/O Devices
}

\author{\IEEEauthorblockN{Botao Yu}
\IEEEauthorblockA{The Ohio State University \\
yu.3737@osu.edu}
\and
\IEEEauthorblockN{Guanqun Song}
\IEEEauthorblockA{The Ohio State University \\
song.2107@osu.edu}
\and
\IEEEauthorblockN{Ting Zhu}
\IEEEauthorblockA{The Ohio State University \\
zhu.3445@osu.edu}
}

\maketitle

\begin{abstract}
Achieving carbon neutrality has become a critical goal in mitigating the environmental impacts of human activities, particularly in the face of global climate challenges. Input/Output (I/O) devices, such as keyboards, mice, displays, and printers, contribute significantly to greenhouse gas emissions through their manufacturing, operation, and disposal processes. In this paper, we explores sustainable strategies for achieving carbon neutrality in I/O devices, emphasizing the importance of environmentally conscious design and development. Through a comprehensive review of existing literature and best approaches, we introduces a framework to outline approaches for reducing the carbon footprint of I/O devices. The result underscore the necessity of integrating sustainability into the lifecycle of I/O devices to support global carbon neutrality goals and promote long-term environmental sustainability.

\end{abstract}


\section{Introduction}

In the face of the global climate problem, the concept of carbon neutrality has gained significant attention as a critical goal for mitigating the impact of human activities on the environment. 
Carbon neutrality refers to the state in which the net greenhouse gas emissions associated with a product, service, or organization are effectively zero, achieved through a combination of emission reduction and offsetting measures \cite{fankhauser2022meaning}. As the world strives to gain a more sustainable future, it is imperative to achieve carbon neutrality in various aspects of human activities.

Input/Output (I/O) devices, such as keyboards, mice, displays, and printers, have significant environmental impacts. The manufacturing processes, used materials, energy consumption during operation, and end-of-life disposal of these devices contribute to greenhouse gas emissions \cite{dawodu2019expert}. Furthermore, the quantity of I/O devices is huge, causing substantial burden to the environment. Therefore, the sustainable design of I/O devices is crucial in reducing their carbon emissions and promoting a more environmentally friendly approach to technology development.

This project report aims to explore the sustainable approaches for I/O devices, with a focus on achieving carbon neutrality. Through a comprehensive literature review and analysis of best practices, this report constructs an approach framework that list various kinds of approach for I/O devices to achieve carbon neutrality.
Additionally, a case study of successful sustainable I/O device designs (Apple Watch Series 9) will be examined to identify key success factors and potential further improvements.

\section{Background Knowledge and Concepts}

\textit{Carbon footprint} is a measure of the total amount of greenhouse gases, primarily carbon dioxide, that are generated by the actions of an individual, organization, event, or product. It is usually expressed in equivalent tons of carbon dioxide (CO2e) and takes into account both direct emissions, such as those from burning fossil fuels, and indirect emissions, which are associated with the full life cycle of goods and services.

\textit{Carbon neutrality} is the concept of achieving a balance between the amount of carbon dioxide emitted into the atmosphere and the amount removed or offset. The goal is to have a net-zero carbon footprint, meaning that the amount of carbon dioxide released is equal to the amount absorbed or eliminated through various methods.

\textit{Carbon credits} are a market-based mechanism designed to reduce greenhouse gas emissions and combat climate change. They represent a unit of measurement that represents the reduction or removal of one metric ton of carbon dioxide (CO2) or its equivalent in other greenhouse gases from the atmosphere. Carbon credits are generated through projects that reduce emissions or enhance carbon sinks, such as renewable energy projects, reforestation efforts, or energy efficiency initiatives. Companies can buy carbon credits to offset their carbon emissions.

\section{Related Work}
Recent years, advancements in wireless networks \cite{wire1,wire3,10017581, 9523755,9340574,10.1145/3387514.3405861,9141221,9120764,10.1145/3356250.3360046,8737525,8694952,10.1145/3274783.3274846,10.1145/3210240.3210346,8486349,8117550,8057109,https://doi.org/10.1155/2017/5156164,10189210}, secure communication \cite{wire2, 10125074,285483,10.1145/3395351.3399367}, smart life \cite{MILLER2022100245,DBLP:journals/corr/abs-2112-15169,YAO2020100087,MILLER2020100089,8556650,10.1145/3127502.3127518,10.1145/3132479.3132480,GAO201718}, and machine learning \cite{10.1145/3460120.3484766,9709070,9444204,ning2021benchmarkingmachinelearningfast,8832180,8556807,8422243,chandrasekaran2022computervisionbasedparking,iqbal2021machinelearningartificialintelligence,pan2020endogenous} 
 have opened new possibilities for addressing the environmental challenges posed by electronic devices. These technologies can optimize energy usage, improve recycling processes through predictive modeling, and enhance the lifecycle management of devices by enabling secure and efficient communication between interconnected systems. This section reviews the current literature on the environmental impact of electronic devices and strategies for achieving sustainable design.

In terms of carbon neutrality, several comprehensive review papers have been published in recent years. \cite{yuan2023carbon} conducted a thorough review of the concept of carbon neutrality, including carbon footprinting, accounting methods, and supporting technologies. They also examined carbon emission reduction technologies and implementation pathways for increasing carbon sequestration. Additionally, they developed a web-based tool for calculating and analyzing the life cycle carbon footprint of products and discussed the practical applications and challenges of digital technologies in achieving carbon neutrality. Another review \cite{chen2022strategies} focused on strategies to achieve a carbon neutral society [2]. They examined the outcome goals of COP 26 and discussed methods for mapping carbon emissions, decarbonization technologies and initiatives, negative emissions technologies, carbon trading, and carbon tax. The authors proposed plans for carbon neutrality, such as shifting towards renewable energy, developing low-carbon technologies and agriculture, changing dietary habits, developing resilient buildings and cities, introducing decentralized energy systems, and electrifying the transportation sector. They also reviewed the life cycle analysis of carbon neutral systems. \cite{wang2021technologies} introduces reviews innovative technologies that offer solutions for achieving carbon neutrality and sustainable development. It includes discussions on renewable energy production, food system transformation, waste valorization, carbon sink conservation, and carbon-negative manufacturing.

Many large tech companies have announced new or revised goals for reaching carbon neutrality. For instance, Apple announced in July 2020 its aim to achieve climate neutrality by 2030, including the whole supply chain and the entire life cycle of Apple products \cite{apple2021environmental}. Other companies such as Google, Microsoft, and Amazon have existing or new commitments, some already claiming carbon neutrality before 2020, and some aiming at the end of 2020. These activities increasingly filter up through the supply chain and now reach electronic packaging as well \cite{EPS2020}.

While the existing literature provides valuable insights into the sustainable design of electronic products and the pursuit of carbon neutrality, there is a lack of comprehensive studies specifically focusing on I/O devices. This project aims to bridge this gap by proposing a holistic framework for sustainable I/O device design, considering various aspects throughout the whole lifecycle of devices.

\section{Approaches of Carbon Neutrality}
\label{sec:approaches}

To achieve carbon neutrality, approaches lie in two aspects: reducing carbon emissions and increasing carbon uptake \cite{yuan2023carbon}. This section introduces these two aspects of approaches in detail, highlighting their significance and potential applications in the context of sustainable I/O device design.

\subsection{Reducing Carbon Emissions}

Reducing carbon emissions is a critical component of achieving carbon neutrality. In the context of I/O device design, this involves implementing strategies that minimize the greenhouse gas emissions associated with the whole lifecycle of devices.
Specifically, the lifecycle of I/O devices can be divided into three phases: production and sales, use, and disposal and recycling. This report will introduce the approaches from the three phases as follows.

\textit{Production and Sales.} A crucial part of carbon emission comes from the production and sales phase, where many raw materials and resources are needed to manufacture those devices, and the shipping also consumes a lot of energy. To reduce carbon emissions in this phase, typical methods are concluded below. 

\paragraph{Eco-Friendly Materials}
One approach to reducing carbon emissions during the production phase is to utilize eco-friendly materials in the manufacturing of I/O devices. This includes using recycled plastics, biodegradable materials, and materials with lower carbon footprints. By incorporating these materials, manufacturers can significantly reduce the environmental impact of I/O device production.

\paragraph{Energy-efficient Manufacturing Processes}
Implementing energy-efficient manufacturing processes is another crucial approach to reducing carbon emissions during production. This can be achieved by optimizing production lines, using renewable energy sources, and adopting lean manufacturing techniques. By minimizing energy consumption and waste during the manufacturing process, companies can substantially reduce their carbon footprint.

\paragraph{Sustainable Packaging and Distribution}
Reducing carbon emissions can also be achieved through sustainable packaging and distribution practices. This involves using recyclable or biodegradable packaging materials, optimizing shipping routes to minimize transportation emissions, and encouraging the use of low-emission vehicles for product distribution. By adopting these practices, companies can further reduce the carbon footprint associated with the production and sales phase.

\textit{Use.} During use, carbon emissions can be reduced by the following methods.

\paragraph{Energy-efficient Designs}
Designing I/O devices with energy efficiency in mind is a key approach to reducing carbon emissions during the use phase. This can be achieved by implementing power management features, such as automatic sleep and hibernation modes, and utilizing low-power components. By reducing the power consumption of I/O devices during operation, manufacturers can help users minimize their carbon footprint.

\paragraph{Software Optimization}
Optimizing device drivers and software can also contribute to reducing carbon emissions during the use phase. Efficient software can help minimize resource consumption, reduce processing overhead, and extend the lifespan of I/O devices. By providing regular software updates and optimizations, manufacturers can ensure that their devices operate at peak efficiency, thereby reducing energy consumption and associated carbon emissions.

\paragraph{User Education and Awareness}
Educating users about sustainable practices and encouraging them to adopt environmentally friendly habits can also help reduce carbon emissions during the use phase. This includes promoting the proper use and maintenance of I/O devices, encouraging users to enable power-saving features, and raising awareness about the environmental impact of their device usage. By empowering users with knowledge and tools to reduce their carbon footprint, manufacturers can foster a more sustainable approach to I/O device use.

\textit{Disposal and Recycling Phase.} At this last phase of lifecycle, devices can be property processed to reduce carbon emissions.

\paragraph{Device Recycling Programs}
Implementing comprehensive device recycling programs is a crucial approach to reducing carbon emissions during the disposal and recycling phase. Manufacturers can establish collection points, collaborate with recycling facilities, and incentivize users to return their old devices for proper recycling. By ensuring that I/O devices are recycled responsibly, companies can minimize the environmental impact of e-waste and conserve resources.

\paragraph{Modular Design for Easy Disassembly}
Designing I/O devices with modularity and easy disassembly in mind can facilitate more efficient recycling and reduce carbon emissions during the disposal phase. By using standardized components and minimizing the use of adhesives and complex fasteners, manufacturers can make it easier for recycling facilities to dismantle and separate different materials. This approach not only simplifies the recycling process but also enables the recovery of valuable materials for reuse.

\paragraph{Extended Producer Responsibility}
Adopting extended producer responsibility (EPR) policies is another approach to reducing carbon emissions during the disposal and recycling phase. EPR holds manufacturers accountable for the environmental impact of their products throughout their lifecycle, including disposal and recycling. By implementing EPR programs, companies can ensure that their I/O devices are properly managed at the end of their life, reducing the burden on municipal waste systems and promoting sustainable disposal practices.

\subsection{Increasing Carbon Uptake}

While reducing carbon emissions is essential, increasing carbon uptake is an equally important strategy for achieving carbon neutrality. Carbon uptake refers to the process of removing carbon dioxide from the atmosphere and storing it in various carbon sinks, such as forests, soils, and oceans. In the context of I/O device design, increasing carbon uptake can be achieved mainly through carbon offsetting.

Carbon offsetting involves investing in projects that remove or prevent carbon dioxide emissions from the atmosphere, effectively compensating for the emissions generated by I/O devices (Fankhauser et al., 2022). These projects can include reforestation initiatives, renewable energy development, and community-based conservation efforts. By supporting these projects, companies can offset the carbon emissions associated with the production and use of their I/O devices, contributing to a net-zero carbon footprint.

Overall, by combining strategies for reducing carbon emissions and increasing carbon uptake, the sustainable design of I/O devices can significantly contribute to the goal of achieving carbon neutrality. A holistic approach that considers the entire lifecycle of these devices, from material selection and manufacturing to use and end-of-life management, is essential for minimizing their environmental impact and promoting a more sustainable future.

\section{Case Study: Apple Watch Series 9}

Apple, a global leader in technology and innovation, has been focusing on promoting sustainable practices in the electronics industry. It has set ambitious targets to reduce its environmental footprint and achieve carbon neutrality across its entire business, including its products, by 2030 \cite{apple2021environmental}. One notable milestone is the Apple Watch Series 9, which has achieved the goal of carbon neutrality \cite{apple2023watch}.

\subsection{Approaches Taken By Apple}

\begin{table*}[htbp]
  \centering
  \caption{Approaches that Apple has taken to achieve carbon neutral on Apple Watch Series 9.}
    \begin{tabular}{ccl}
    \toprule
    \textbf{Phase} & \textbf{Aspect} & \textbf{Apple's approach} \\
    \midrule
    \rowcolor[rgb]{ .792,  .929,  .984} \multicolumn{3}{c}{Reducing Carbon Emissions} \\
    \midrule
    \multirow{14}[6]{*}{Production and Sales} & \multirow{8}[2]{*}{Eco-Friendly Materials} & 100\% recycled aluminum for the case \\
          &       & 100\% recycled rare earth elements in magnets (99\% of total rare earths) \\
          &       & 100\% recycled copper foil in main logic board and copper wire in Taptic Engine \\
          &       & 25\% recycled plastic in multiple components, renewable plastic used in speaker \\
          &       & 100\% recycled tungsten in Taptic Engine (100\% of total tungsten) \\
          &       & 100\% recycled cobalt in battery \\
          &       & 100\% recycled tin in solder of multiple PCBs \\
          &       & 100\% recycled gold in plating of multiple PCBs \\
\cmidrule{2-3}          & \multirow{2}[2]{*}{Energy-efficient Manufacturing Processes} & 100\% of manufacturing electricity sourced from clean energy \\
          &       & Apple suppliers committed to Apple's Supplier Clean Energy Program \\
\cmidrule{2-3}          & \multirow{4}[2]{*}{Sustainable Packaging and Distribution} & 100\% fiber-based packaging, working to eliminate plastic \\
          &       & 47\% recycled content in fiber packaging \\
          &       & 100\% of new wood fiber from responsibly managed forests \\
          &       & 50\% or more products shipped via non-air freight \\
    \midrule
    \multirow{4}[6]{*}{Use} & \multirow{2}[2]{*}{Energy-efficient Designs} & Meets California's energy efficiency requirements \\
          &       & Uses power-efficient components that manage power consumption \\
\cmidrule{2-3}          & Software Optimization & Software intelligently manages power \\
\cmidrule{2-3}          & User Education and Awareness & Engaging customers to support grid decarbonization \\
    \midrule
    \multirow{5}[6]{*}{Disposal and Recycling} & \multirow{3}[2]{*}{Device Recycling Programs} & Apple Trade In program to return old devices for recycling \\
          &       & Holds recyclers to high standards \\
          &       & Provides recycler guides for safe disassembly and material recovery \\
\cmidrule{2-3}          & Modular Design for Easy Disassembly & \textcolor[rgb]{ 1,  0,  0}{Not applicable} \\
\cmidrule{2-3}          & Extended Producer Responsibility & Offers take-back and recycling in 99\% of countries products are sold \\
    \midrule
    \rowcolor[rgb]{ .792,  .929,  .984} \multicolumn{3}{c}{Increasing Carbon Uptake} \\
    \midrule
    \multirow{2}[2]{*}{-} & \multirow{2}[2]{*}{Carbon Offsetting} & Invests in nature-based carbon removal projects like the Restore Fund \\
          &       & Carefully selects high-quality avoided deforestation and carbon removal offset projects \\
    \bottomrule
    \end{tabular}%
  \label{tab:aw_approaches}%
\end{table*}%

The approaches that Apple has taken on Apple Watch Series 9 is listed in \cref{tab:aw_approaches}, which have been summarized from the aspects mentioned in \cref{sec:approaches}.

\textit{Reducing Carbon Emissions.}
Apple has taken significant steps to reduce carbon emissions throughout the life cycle of the Apple Watch Series 9. 

In the production and sales phase, they have focused on using eco-friendly materials, such as 100\% recycled aluminum, rare earth elements, copper, tungsten, cobalt, tin, and gold in various components. Apple has also ensured that 100\% of the manufacturing electricity is sourced from clean energy and has engaged suppliers in their Clean Energy Program. Additionally, they have prioritized sustainable packaging and distribution, using 100\% fiber-based packaging with 47\% recycled content, responsibly sourced wood fiber, and shipping 50\% or more products via non-air freight.

During the use phase, Apple has designed the Watch to be energy-efficient, meeting California's stringent requirements and utilizing power-efficient components and intelligent software to manage power consumption. They also engage with customers to support grid decarbonization.

In terms of disposal and recycling, Apple offers the Trade In program to facilitate the return of old devices for recycling, holds recyclers to high standards, and provides guides for safe disassembly and material recovery. The company also offers take-back and recycling services in 99\% of the countries where their products are sold.

The Apple Watch Series 9 serves as a compelling case study for the sustainable design of I/O devices and the pursuit of carbon neutrality in the technology industry. By incorporating eco-friendly materials, optimizing energy efficiency, minimizing packaging waste, promoting end-of-life management, and investing in carbon offsetting projects, Apple has demonstrated the feasibility of creating a carbon-neutral I/O device. This case study highlights the importance of a holistic approach to sustainability, considering the entire lifecycle of the product and the broader environmental impact of its production, use, and disposal.

\textit{Increasing Carbon Uptake.}
To offset the remaining carbon emissions, Apple invests in nature-based carbon removal projects like the Restore Fund. They carefully select high-quality avoided deforestation and carbon removal offset projects to ensure that the carbon uptake is real, additional, and permanent.

By combining these approaches to reduce emissions and increase carbon uptake, Apple has achieved carbon neutrality for the Apple Watch Series 9 paired with the Sport Loop band.

\subsection{Analysis and Comments}

\begin{figure}
    \centering
    \includegraphics[width=0.45\textwidth]{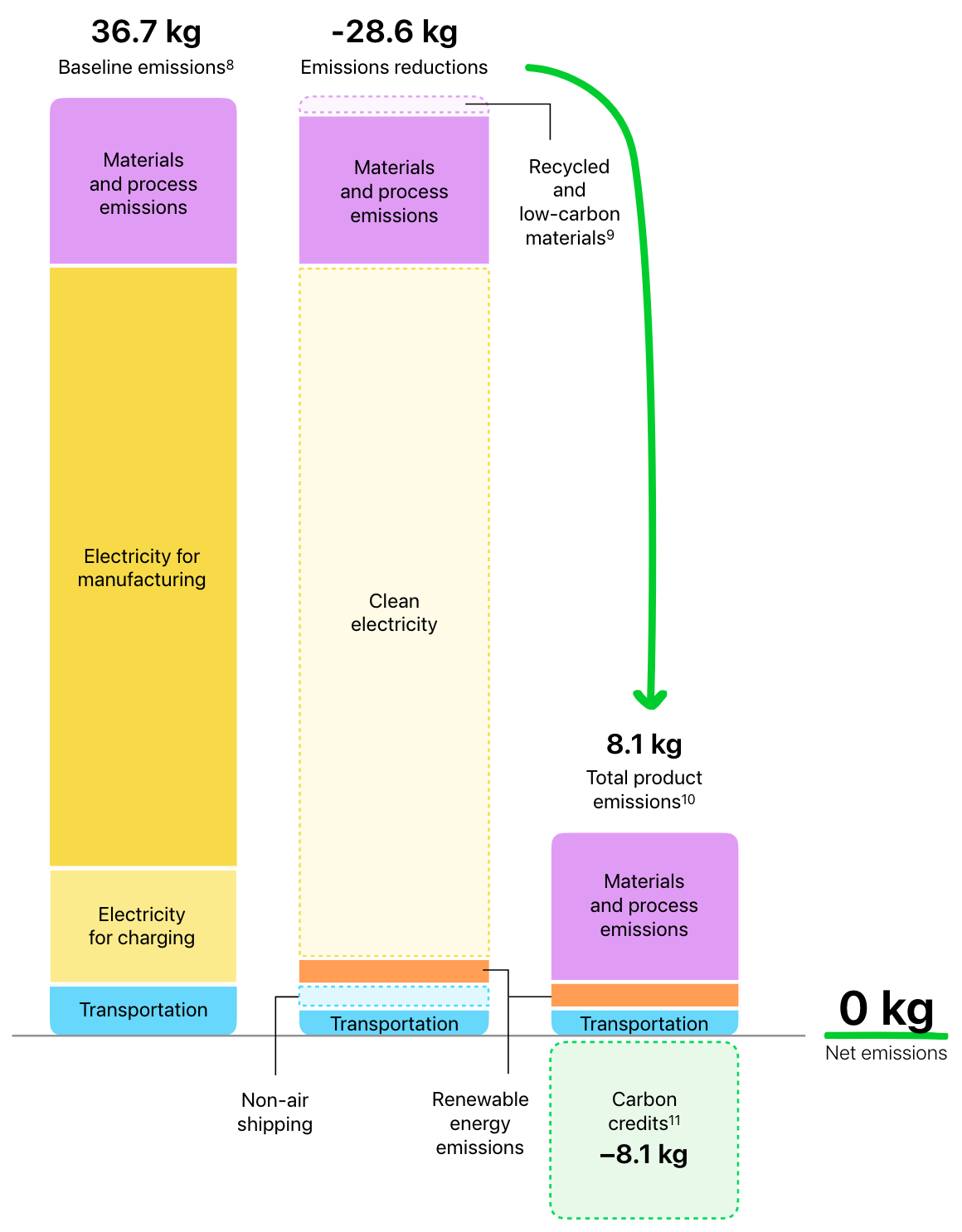}
    \caption{Carbon neutral of Apple Watch Series 9 \cite{apple2023watch}.}
    \label{fig:aw_cd}
\end{figure}

\cref{fig:aw_cd} illustrates how Apple has reduced the carbon emissions for the Apple Watch Series 9 paired with the Sport Loop band. The baseline emissions were 36.7 kg, but through the use of recycled and low-carbon materials, clean electricity for manufacturing, non-air shipping, and renewable energy emissions, Apple has achieved a reduction of 28.6 kg in emissions.
The total product emissions now stand at 8.1 kg, which includes materials and process emissions, as well as transportation. To offset these remaining emissions, Apple has invested in carbon credits amounting to 8.1 kg, resulting in net zero emissions for the product.

While Apple's efforts to achieve carbon neutrality for the Apple Watch Series 9 and Sport Loop are commendable, there are some areas where their approaches could be further improved:

\paragraph{Reliance on carbon offsets}
Although Apple has made significant strides in reducing emissions through the use of recycled materials, clean energy, and efficiency improvements, they still rely on purchasing carbon credits to offset the remaining 8.1 kg of emissions. While carbon offsets can be a useful tool, they should be seen as a last resort after all possible emission reductions have been made. Apple should continue to focus on further reducing emissions through innovation and efficiency gains, rather than relying too heavily on offsets.

\paragraph{Transparency in carbon offset projects}
While Apple states that they carefully select high-quality carbon offset projects, there is limited information provided about the specific projects they invest in. Increased transparency about these projects, including their location, type, and verification process, would help to build trust in Apple's carbon neutrality claims.

\paragraph{Addressing other environmental impacts}
Although reducing carbon emissions is crucial, it is not the only environmental challenge facing the electronics industry. Apple should also prioritize addressing other impacts, such as resource depletion, waste generation, and the use of hazardous chemicals in their products and supply chain.

\paragraph{Encouraging longer product lifespans}
One of the most effective ways to reduce the environmental impact of electronics is to extend their useful lifespan. While Apple does offer some repair and recycling services, they could do more to design their products for longevity, repairability, and upgradeability, which would help to reduce the need for frequent replacements and the associated carbon emissions.

\section{Encountered Issues}

During the research process for this project, the primary issues encountered during the research process was the limited availability and consistency of data related to the environmental impact of I/O devices. While there are numerous studies on achieving carbon neutrality in general, specific information sources on I/O devices are often scarce. This lack of comprehensive and consistent information made it challenging to quantify the carbon footprint of I/O devices accurately and to compare the environmental performance of different strategies.

\section{Conclusion}

This project report has explored various approaches to achieving carbon neutrality in the sustainable design of I/O devices, focusing on strategies for reducing carbon emissions and increasing carbon uptake throughout the product lifecycle. The case study of the Apple Watch Series 9 demonstrates the successful implementation of these strategies, while also highlighting areas for improvement.

Limitations of this study include the lack of quantitative analysis of the environmental impact of specific strategies and the focus on a single case study. Future research could expand on this work by conducting more in-depth analyses of various I/O devices and their lifecycle impacts, as well as exploring the effectiveness of different sustainability approaches in a wider range of contexts.

Moving forward, it is essential for the industry to prioritize sustainability in I/O device design, collaborate on establishing standardized metrics and reporting practices, and invest in research and development to drive further innovation in low-carbon technologies. By working together and sharing best practices, the technology sector can play a vital role in mitigating the impact of human activities on the environment and ensuring a more sustainable future for generations to come.


\bibliography{reference,zhu}
\bibliographystyle{ieeetr}

\end{document}